\documentclass[aps,prd,twocolumn,showpacs,nofootinbib]{revtex4}

\usepackage{amsfonts,color,graphicx,epsfig,amsmath,multirow,slashed}

\usepackage[colorlinks=true,linkcolor=blue,citecolor=blue, urlcolor=blue]{hyperref}

\begin{document}

\title{The prompt $J/\psi$ production in Association with a $c\bar{c}$ Pair within the Framework of Non-relativistic QCD via photon-photon collision at the International Linear Collider}

\author{Zhan Sun}
\email{zhansun@cqu.edu.cn}

\author{Xing-Gang Wu}
\email{wuxg@cqu.edu.cn}
\address{Department of Physics, Chongqing University, Chongqing 401331, P.R. China}
\address{Institute of Theoretical Physics, Chongqing University, Chongqing 401331, P.R. China}

\author{Hong-Fei Zhang}
\email{hfzhang@ihep.ac.cn}
\address{Department of Physics, School of Biomedical Engineering, Third Military Medical University, Chongqing 400038, P.R. China}
\address{Chongqing University of Posts $\&$ Telecommunications, Chongqing, 400065, P.R. China}

\date{\today}

\begin{abstract}

We present a systematical study on the prompt $J/\psi$ production in association with a $c\bar{c}$ pair via the process, $\gamma\gamma \to H(c\bar{c}) +c+\bar{c}$, within the framework of non-relativistic QCD at the future high-energy $e^{+}e^{-}$ collider - International Linear Collider (ILC), including both direct and feed-down contributions. For direct $J/\psi$ production, the states with color-octet channels, especially the $^3P_J^{[8]}$ and $^1S_0^{[8]}$ ones, provide dominant contribution to the production cross-section, which are about fifty-two times over that of the color-singlet one. This is clearly shown by the transverse momentum ($p_t$) and rapidity distributions. The feed-down contribution from $\psi'$ and $\chi_{cJ}$ ($J=0,1,2$) is sizable, which is about $20\%$ to the total prompt cross-section. Besides the yields, we also calculate the $J/\psi$ polarization parameter $\lambda$. In small $p_t$ region, the polarization of the prompt $J/\psi$ is longitudinal due to large contributions through the $^3P_J^{[8]}$ channel, and becomes transverse in high $p_t$ region due to $^3S_1^{[8]}$ channel. Thus the $J/\psi$ production via photon-photon collisions at the ILC shall provide a useful platform for testing the color-octet mechanism.

\end{abstract}


\maketitle

\section{Introduction}

With its heavy mass $m$ the heavy quark ($Q$) will move with a small velocity $v_Q$ inside a heavy quarkonium. This results in a hierarchy of energy scales, $m >> m v_Q >> m v_Q^2$, and the dynamics at different energy scales is different. The non-relativistic QCD (NRQCD) provides a systematical way to separate the effects from the dynamics at different energy scales~\cite{Bodwin:1994jh}. Since its invention, it has been widely applied to deal with heavy quarkonium physics. It provides a solution to the $\psi'$ ($J/\psi$) surplus puzzle~\cite{Braaten:1994vv, Cho:1995vh, Cho:1995ce}. It gives a consistent explanation for both the $J/\psi$ polarization and the $\eta_c$ production data at hadron colliders~\cite{Aaij:2013nlm, Chatrchyan:2013cla, Aaij:2014bga} by using the newly fitted color-octet long distance matrix elements (LDMEs)~\cite{Zhang:2014ybe, Sun:2015pia}. It gives an explanation for the $\chi_{cJ}$-production data measured by the hadronic colliders~\cite{Jia:2014jfa}. A comprehensive review of NRQCD theory and applications can be found in Refs.~\cite{Brambilla:2010cs, Brambilla:2014jmp}.

The NRQCD theory still faces many challenges. As an important example, the Belle and the LHCb collaborations~\cite{Abe:2002rb, Aaij:2012dz} have measured the $J/\psi$ production associated with a $c\bar{c}$ pair, which however show large discrepancies from the NRQCD predictions~\cite{Liu:2003jj, Hagiwara:2004pf, Zhang:2006ay, Gong:2009ng, He:2007te, Berezhnoy:1998aa, Baranov:2006dh, Lansberg:2008gk}. For instance, the predicted angular distribution of $J/\psi$ is different from the Belle data, and the predicted $J/\psi$ production cross-section is lower than the LHCb data by about one order of magnitude. This discrepancy may indicate a breaking of NRQCD factorization in small $J/\psi$ momentum region at the Belle. Thus it is helpful to find more experimental platforms to test NRQCD, especially those with high collision energies.

A detailed prediction for the $J/\psi$ production associated with a $c\bar{c}$ pair, $e^+ e^-\to J/\psi+c\bar{c}$, at the super-$Z$ factory has been done in Ref.\cite{Sun:2013wuk}. Due to the $Z^0$-boson resonance effect, sizable $J/\psi$ events can be achieved there. The proposed International Linear Collider (ILC)~\cite{Djouadi:2007ik} is another useful $e^+ e^-$ collider, which has been designed to run at a high collision energy from several hundred GeV to TeV together with a high luminosity about ${\cal L} \simeq 10^{34-36}{\rm cm}^{-2} {\rm s}^{-1}$. An analysis of $J/\psi$ production in association with a $c\bar{c}$ pair via photon-photon collision at the ILC has been done by Ref.~\cite{Chen:2014xka}. It shows sizable $J/\psi$-events, about $4.7 \times 10^{4}$ per year, can be generated under the condition of $\sqrt{S}=500$ GeV and ${\cal L}\simeq 10^{34}$cm$^{-2}$s$^{-1}$. As a comparison, the $J/\psi$ photon-photon production in association with light hadrons is very small~\cite{Ma:1997bi} \footnote{It only contributes $0.4\%$ and $0.02\%$ to the $J/\psi$ inclusive production cross-section with $\sqrt{S}=500$ GeV and $1$ TeV, respectively.}.

It is noted that in Ref.\cite{Chen:2014xka}, the $J/\psi$ is generated via the direct production channel, $\gamma\gamma \to J/\psi+c\bar{c}$, in which only the color-singlet contribution has been taken into consideration. According to our NRQCD practices, the contributions from the color-octet charmonium states may be sizable or even dominant. Furthermore, the feed-down contributions from the $\psi'$ and $\chi_{cJ}$ may also be sizable, i.e. one can first generate the $\psi'$ and $\chi_{cJ}$ mesons via the channel, $e^+ e^-\to \psi'/\chi_{cJ} +c\bar{c}$, and then they may decay to $J/\psi$ via the channels $\psi'\to J/\psi+X$, $\chi_{cJ}\to J/\psi+\gamma$ and $\psi'\to\chi_{cJ}\to J/\psi+\gamma$. Thus, as a sound a NRQCD prediction, we shall in this paper study the prompt $J/\psi$ production with a $c\bar{c}$ pair, including both the direct and the feed-down contributions from $\psi'$ and $\chi_{cJ}$ mesons.

For the direct $J/\psi$ and $\psi'$ production, four intermediate states shall be considered, i.e. the color-singlet one ($^3S_1^{[1]}$) and the three color-octet ones ($^1S_0^{[8]}$, $^3S_1^{[8]}$ and $^3P_J^{[8]}$), which keep the NRQCD expansion up to the order of $v_Q^4$. For the direct $\chi_{cJ}$ production, up to the leading order in $v_Q$, only the two dominant states ($^3P_J^{[1]}$ and $^3S_1^{[8]}$) shall be considered. The feeddown contributions from $\psi'$ to $\chi_{cJ}$ shall also be considered in our work, which, as shall be shown, give $\sim 2\%$ contribution to the prompt $J/\psi$ production.

The remaining parts of the paper are organized as follows. In Sec.II, we describe the calculation technology for treating the $J/\psi$ production in association with a $c\bar{c}$-pair. The input parameters are also introduced there. In Sec.III, we give the predictions for the yields and polarizations of $J/\psi$ at the ILC. The last section is reserved for a summary.

\section{Calculation Technology and Input Parameters}

At the ILC, the charmonium $H$ can be produced via the process $e^{+} + e^{-} \to e^{+} + e^{-} + H(c\bar{c}) +c+\bar{c}$. For the photon-photon production, $\gamma\gamma \to H(c\bar{c}) +c+\bar{c}$, the photons may interact either directly (direct photon production) or indirectly via their hadronic components (resolved photon production) with the quarks or anti-quarks. In the paper we shall focus our attention on leading contribution from the direction photon production. Within the NRQCD framework, the differential cross-section for the charmonium $H$ can be factorized as
\begin{eqnarray}
d\sigma^{H} &=&\sum_{n}\int dx_1 dx_2 f_{\gamma}(x_1)f_{\gamma}(x_2) \nonumber \\
&&~~~~~~\times {d\hat{\sigma}(\gamma\gamma\to c\bar{c}[n]+c+\bar{c})}\langle \mathcal O ^{H}(n) \rangle.
\end{eqnarray}
$\hat{\sigma}$ stands for short-distance cross-section, representing the production of an intermediate perturbative state $(c\bar{c})[n]$ with quantum number $n$. $\langle \mathcal{O}^{H}(n) \rangle$ is non-perturbative but universal LDMEs, which cannot be evaluated perturbatively, but can be obtained through the fit of the experiment. $f_{\gamma}(x)$ is the photon density function with $x$ being the momentum fraction of the photon to the initial electron or positron.

\begin{figure}[htb]
\begin{center}
\hspace{0cm}\includegraphics[width=0.48\textwidth]{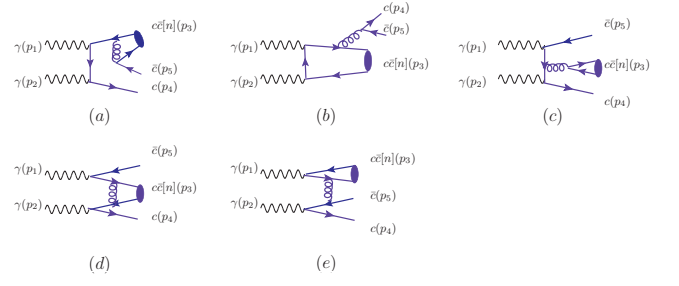}
\caption{Typical Feynman Diagrams for the hard subprocess $\gamma(p_1)\gamma(p_2) \to c\bar{c}[n](p_3)+c(p_4)+\bar{c}(p_t)$, in which ($a,d$) are for color-singlet states, ($a,b,c,d$) are for $c\bar{c}[^3S_1^{[8]}]$ and ($a,d,e$) are for $c\bar{c}[^1S_0^{[8]}]$ and $c\bar{c}[^3P_J^{[8]}]$, respectively.} \label{FD}
\end{center}
\end{figure}

For the hard subprocess $\gamma\gamma \to c\bar{c}[n]+c+\bar{c}$, there are totally $20$ diagrams for $[n]={^3S_1^{[1]}}$ and $[n]= {^3P_J^{[1]}}$, $32$ diagrams for $[n]= {^3S_1^{[8]}}$, and $28$ diagrams for $[n]= {^1S_0^{[8]}}$ and $[n]= {^3P_J^{[8]}}$. Typical Feynman diagrams are presented in Fig.(\ref{FD}). Those Feynman diagrams can be generated and the corresponding short-distance cross-sections can be calculated by using the well-established FDC package. The FDC is a general-purpose program for Feynman diagram calculation which realizes an automatic deduction from physical models to final numerical results for lower-order processes~\cite{Wang:2004du}.

At the high-energy $e^+ e^-$ collider such as ILC, the laser backscattering (LBS) from the incident electron and positron beams leads to high luminosity photon beams~\cite{ydenfun}. Those LBS photons are hard enough and carry a large fraction of the energy of the incident lepton beams. Moreover, as pointed out by Refs.~\cite{Ma:1997bi, Li:2009fd, Qiao:2003ba}, using LBS photons the $\gamma\gamma$ collisions can  approximately achieve the same high luminosity as that of the $e^+e^-$ beams. Thus, we adopt the LBS density function to do our calculation~\cite{ydenfun}
\begin{eqnarray}
f_{\gamma}(x) &=& \frac{1}{N} \left[1-x+\frac{1}{1-x}-4r(1-r)\right] ,
\end{eqnarray}
where $r=x/[x_m(1-x)]$ and the normalization factor
\begin{eqnarray}
N &=& \left(1-\frac{4}{x_m}-\frac{8}{x_m^2}\right)\log\chi +\frac{1}{2}+\frac{8}{x_m} -\frac{1}{2\chi^2}.
\end{eqnarray}
Here $\chi=1+x_m$, $x_m \simeq 4.83$~\cite{Telnov:1989sd} and the LBS photon energy is restricted by $0 \leq x \leq {x_m}/{(1+x_m)}=0.83$.

The unpolarized differential cross-section of prompt $J/\psi$ is obtained by adding the unpolarized cross-sections of the various direct-production processes multiplied by appropriate branching fractions, i.e.
\begin{eqnarray}
&& d\sigma^{\rm prompt\; J/\psi} \nonumber\\
&=& d\sigma^{J/\psi} + \sum_{J} \mathcal{B}r(\chi_{cJ}\to J/\psi +\gamma)\; d\sigma^{\chi_{cJ}} + \nonumber\\
&& \sum_{J}\mathcal{B}r(\psi'\to \chi_{cJ} +\gamma) \mathcal{B}r(\chi_{cJ}\to J/\psi +\gamma)\; d\sigma^{\psi'} \nonumber\\
&& + \mathcal{B}r(\psi'\to J/\psi +X)\; d\sigma^{\psi'}.
\end{eqnarray}
As for the branching ratios, we take~\cite{Agashe:2014kda}
\begin{eqnarray}
\mathcal{B}r(\psi^{\prime} \to J/\psi+X)&=&60.9\%, \nonumber \\
\mathcal{B}r(\psi^{\prime} \to \chi_{c0}+\gamma)&=&9.99\%, \nonumber \\
\mathcal{B}r(\psi^{\prime} \to \chi_{c1}+\gamma)&=&9.55\%, \nonumber \\
\mathcal{B}r(\psi^{\prime} \to \chi_{c2}+\gamma)&=&9.11\%, \nonumber \\
\mathcal{B}r(\chi_{c0} \to J/\psi+\gamma)&=&1.27\%, \nonumber \\
\mathcal{B}r(\chi_{c1} \to J/\psi+\gamma)&=&33.9\%, \nonumber \\
\mathcal{B}r(\chi_{c2} \to J/\psi+\gamma)&=&19.2\%. \nonumber
\end{eqnarray}

As for the LDMEs, we take~\cite{Jia:2014jfa, Zhang:2014ybe, Sun:2015pia}:
\begin{eqnarray}
\langle O^{\chi_{cJ}}(^3S_1^{[8]})\rangle&=&(2J+1)\times 2.15 \times 10^{-3} ~\rm{GeV^3}, \nonumber \\
\langle O^{\chi_{cJ}}(^3P_J^{[1]})\rangle &=&(2J+1)\times\frac{3}{4\pi}|R_{1P}^{\prime}(0)|^2, \nonumber \\
\langle O^{\psi^{\prime}}(^1S_0^{[8]})\rangle&=&1.0 \times 10^{-2}~\rm{GeV^3}, \nonumber \\
\langle O^{\psi^{\prime}}(^3S_1^{[8]})\rangle&=&0.401 \times 10^{-2}~\rm{GeV^3}, \nonumber \\
\frac{\langle O^{\psi^{\prime}}(^3P_0^{[8]})\rangle}{m^2_c}&=&0.682 \times 10^{-2}~\rm{GeV^3}, \nonumber \\
\langle O^{\psi^{\prime}}(^3S_1^{[1]})\rangle &=&\frac{1}{4\pi}|R_{2S}(0)|^2, \nonumber \\
\langle O^{J/\psi}(^3S_1^{[1]})\rangle&=&0.645~\rm{GeV^3}, \nonumber \\
\langle O^{J/\psi}(^1S_0^{[8]})\rangle&=&0.780 \times 10^{-2}~\rm{GeV^3}, \nonumber \\
\langle O^{J/\psi}(^3S_1^{[8]})\rangle&=&1.057 \times 10^{-2}~\rm{GeV^3}, \nonumber \\
\frac{\langle O^{J/\psi}(^3P_0^{[8]})\rangle}{m^2_c}&=&1.934 \times 10^{-2}~\rm{GeV^3}, \nonumber
\end{eqnarray}
where $J=(0,1,2)$, the squared first derivative of the radial wave-function at the origin $|R_{1P}^{\prime}(0)|^2 =0.075~\rm{GeV}^5$ and the squared radial wave-function at the origin $|R_{2S}(0)|^2 =0.529~\rm{GeV}^3$~\cite{Eichten:1995ch}. We have taken the central value of $\langle O^{J/\psi}(^3S_1^{[1]})\rangle$ to be the one determined from a recently determined $\langle O^{\eta_c}(^1S_0^{[n]})\rangle$~\cite{Zhang:2014ybe}, which is different from the one extracted from measurements of $\Gamma(J/\psi \to l^+l^-)$. In Ref.\cite{Zhang:2014ybe}, the color-singlet LDME for $\eta_c$ has been determined via a direct fitting of the new LHCb data on the $\eta_c$ production~\cite{Aaij:2014bga}, which gives, $\langle O^{\eta_c}(^1S_0^{[1]})\rangle=0.215\pm0.135~\textrm{GeV}^{3}$. This value is comparable with the ones obtained by other groups, e.g., $0.39~\textrm{GeV}^3$~\cite{Eichten:1995ch} and  $0.437^{+0.111}_{-0.105}~\textrm{GeV}^{3}$~\cite{Bodwin:2007fz}. By further using the heavy-quark spin symmetry, $\langle O^{\eta_c}(^1S_0^{[n]})\rangle =\frac{1}{3}\langle O^{J/\psi}(^3S_1^{[n]})\rangle$, we obtain $\langle O^{J/\psi}(^3S_1^{[1]})\rangle=0.645\pm0.405~\textrm{GeV}^3$.

\section{Numerical Results and Discussions}

To do the numerical calculation, we take $m_c=1.5$ GeV and $\alpha=1/137$. The one-loop $\alpha_s$ running is used, and for each charmonium $H$, we set its renormalization and factorization scale as $\mu_{R}=\mu_{F}=M_t=\sqrt{4m_c^2+(p^{H}_t)^2}$ \footnote{Taking $\mu_R=M_t/2$, the prompt cross-section shall be increased by $70\%$, and taking $\mu_R=2M_t$, the prompt cross-section shall be decreased by $30\%$. Such a large scale error could be suppressed by a high-order calculation and/or a proper scale setting~\cite{Wu:2013ei}.}. Regarding the feed-down contributions, the resultant $J/\psi$ transverse momentum can be achieved by a shift, $p^{J/\psi}_t = \frac{m_{J/\psi}}{m_{H}}p^{H}_t$, where $m_{H}$ stands for charmonium mass. The charmonium masses are~\cite{Agashe:2014kda}: $m_{J/\psi}=3.097~\rm{GeV}$, $m_{\psi^{\prime}}=3.686~\rm{GeV}$, $m_{\chi_{c0}}=3.415~\rm{GeV}$, $m_{\chi_{c1}}=3.511~\rm{GeV}$, and $m_{\chi_{c2}}=3.556~\rm{GeV}$.

\begin{table}[htb]
\begin{tabular}{c c c c c}
\hline
 ~~$\sqrt{S}$~~ & ~~$\sigma^{J/\psi}_{^3S^{[1]}_1}$~~ & ~~$\sigma^{J/\psi}_{^1S^{[8]}_0}$~~ & ~~$\sigma^{J/\psi}_{^3S^{[8]}_1}$~~ & ~~$\sigma^{J/\psi}_{^3P^{[8]}_J}$\\
\hline
$250~\textrm{GeV}$ & $9.28 \times 10^{-4}$ & $1.40 \times 10^{-3}$ & $3.46 \times 10^{-5}$ & $1.94 \times 10^{-2}$\\
$500~\textrm{GeV}$ & $4.03 \times 10^{-4}$ & $1.09 \times 10^{-3}$ & $1.62 \times 10^{-5}$ & $1.46 \times 10^{-2}$\\
$1000~\textrm{GeV}$ & $1.58 \times 10^{-4}$ & $5.66 \times 10^{-4}$ & $6.87 \times 10^{-6}$ & $7.60 \times 10^{-3}$\\
\hline
\end{tabular}
\caption{The integrated cross-section (in unit: nb) for the direct $J/\psi$ photon-photon production in association with a $c\bar{c}$ pair at the ILC. $|y|<4.5$. } \label{directcrossection}
\end{table}

\begin{table}[htb]
\begin{tabular}{c c c c c}
\hline
 ~~$\sqrt{S}$~~ & ~~$\sigma_{\textrm{direct}}$~~ & ~~$\sigma_{\textrm{feed-down}~\textrm{from}~\psi^{\prime}}$ & ~~$\sigma_{\textrm{feed-down}~\textrm{from}~\chi_{cJ}}$\\
\hline
$250~\textrm{GeV}$ & $2.18 \times 10^{-2}$ & $5.79 \times 10^{-3}$ & $6.29 \times 10^{-4}$\\
$500~\textrm{GeV}$ & $1.61 \times 10^{-2}$ & $4.18 \times 10^{-3}$ & $4.16 \times 10^{-4}$\\
$1000~\textrm{GeV}$ & $8.33 \times 10^{-3}$ & $2.14 \times 10^{-3}$ & $2.07 \times 10^{-4}$\\
\hline
\end{tabular}
\caption{The integrated cross-section (in unit: nb) for the prompt $J/\psi$ photon-photon production in association with a $c\bar{c}$ pair at the ILC. As for the direct $J/\psi$ production, the contributions through the four channels have been summed up. $|y|<4.5$. }\label{promptcrossection}
\end{table}

We present the integrated cross-section for the $J/\psi$ photon-photon production in association with a $c\bar{c}$ pair in Tables \ref{directcrossection} and \ref{promptcrossection}. The integrated cross-section decreases with the increment of the $e^+ e^-$ collision energy $\sqrt{S}$. The photon-photon production of $J/\psi$ is dominated by direct production, while the feed-down contribution is sizable. More explicitly, when $\sqrt{S}=1$ TeV, the direct channel provide $\sim78\%$ contribution to the prompt $J/\psi$ production, that of the feed-down from $\psi^{\prime}$ is $\sim 20\%$, and that of feed-down from $\chi_{cJ}$ is $\sim 2\%$. Furthermore, for the direct production, the color-octet contributions are large; two channels ($^3P^{[8]}_J$ and $^1S^{[8]}_0$) provide over $95\%$ contribution to the direct $J/\psi$ cross-section. Thus the future new measurements on $J/\psi$ photoproduction at the ILC can provide a useful platform for testing the NRQCD color-octet mechanism.

As an important point, we provide an explanation of the reason why $^1S^{[8]}_0$ and $^3P^{[8]}_J$ channels provide such large contributions in comparison with the other two. By analyzing the topologies of Feynman diagrams, it is noted that the largest contributions of $^1S^{[8]}_0$ and $^3P^{[8]}_J$ mainly come from Fig.(\ref{FD}.(e)), which is absent in the other two channels due to the C-parity and color conservation. The squared invariant mass of the internal gluon of Fig.(\ref{FD}.(e)) is, $k^2=(p_1-p_3)^2=4m^2_c-2 p_1 \cdot p_3$ with $p_1 \cdot p_3=x \frac{\sqrt{S}}{2}(E_{H}-p_{3z})$, where $E_{H}$ (=$\frac{e^y+e^{-y}}{2} M_t$) is the charmonium $H$ energy, $p_{3z}$ (=$\frac{e^y-e^{-y}}{2}M_t$) is the projection of its 3-momentum to the flying direction of the initial photon which attaches to $H$, and $y$ is the rapidity of $H$. $H$ stands for $^1S^{[8]}_0$ or $^3P^{[8]}_J$, and $x$ is the momentum fraction of this photon to the electron or positron. Then the $k^2$-expression can be rewritten as
\begin{eqnarray}
k^2 &=& 4m^2_c - x \sqrt{S} M_t e^{-y} \nonumber \\
&=& 4m^2_c-2(E_{H}+E_{g})\times {M_t}e^{-y} \nonumber \\
&=& -4m^2_c e^{-2y}-(1+e^{-2y})(p^H_t)^2-2E_{g}{M_t}e^{-y}, \label{gluonmass}
\end{eqnarray}
where $E_g$ is the energy of the internal gluon. In small $p_t$ region, the magnitude of $k^2$ reduces exponentially with the increment of $y$ and could be very small, leading to large contributions to the production cross-section.

\begin{figure}[tb]
\begin{center}
\includegraphics[width=0.48\textwidth]{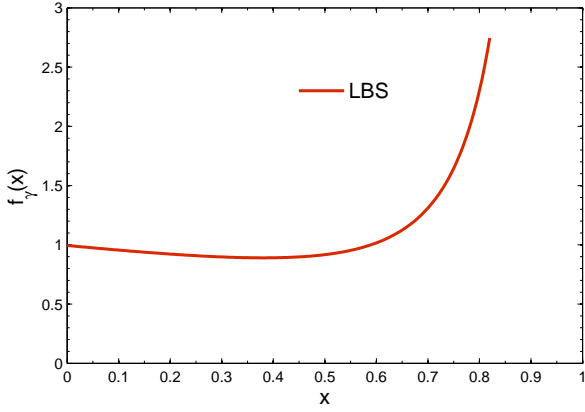}
\caption{The LBS photon density function.} \label{pdf}
\end{center}
\end{figure}

In addition to the topology, the LBS photon density function provides another important factor for large production cross-sections. A bigger energy $E_{H}$ corresponds to a bigger rapidity $|y|$, and vice versa. On the other hand, a larger photon momentum fraction $x$ indicates more $H$ events with larger energy can be achieved. As shown by Fig.(\ref{pdf}), the LBS photon density function behaves moderately in small and intermediate $x$ regions and increases fast when approaching to its allowable largest $x$. Thus the contributions of Fig.(\ref{FD}.(e)) will be further enhanced by LBS photons.

\begin{figure}[tb]
\begin{center}
\includegraphics[width=0.48\textwidth]{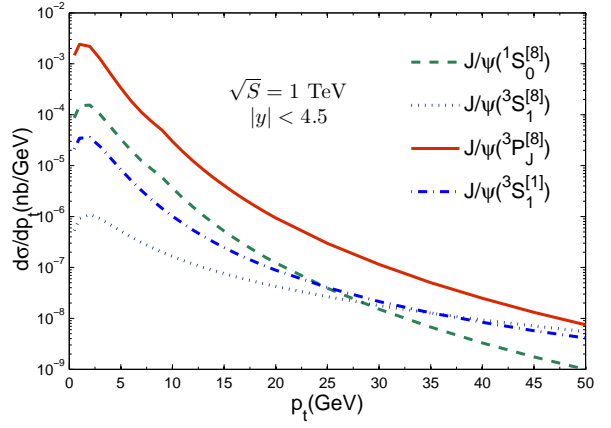}
\caption{The $p_t$-distributions for the direct $J/\psi$ production associated with a $c\bar{c}$ pair at the ILC. The dash-dot, the dashed, the dotted and the solid lines are for $^3S^{[1]}_1$, $^1S^{[8]}_0$, $^3S^{[8]}_1$ and $^3P^{[8]}_J$, respectively.} \label{jpsiptdirect}
\end{center}
\end{figure}

\begin{figure}[tb]
\begin{center}
\includegraphics[width=0.48\textwidth]{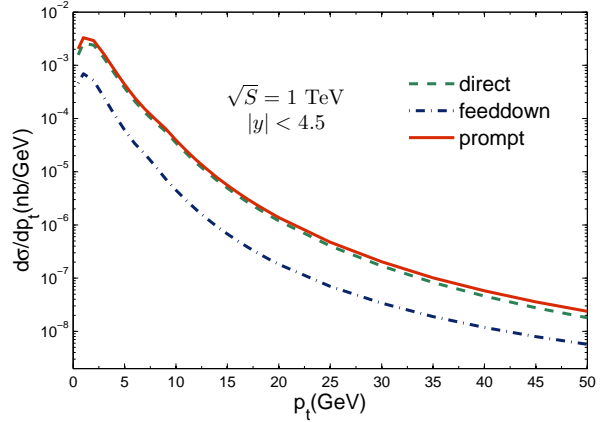}
\caption{The $p_t$-distributions for the prompt $J/\psi$ production associated with a $c\bar{c}$ pair at the ILC. The dashed line is for direct $J/\psi$ production, in which the contributions through the four channels have been summed up. The dash-dot line is for the feed-down contribution from $\psi^{\prime}$ and $\chi_{cJ}$ decay. The solid line is for the prompt production.} \label{jpsiptprompt}
\end{center}
\end{figure}

We present the $p_t$-distributions of the direct and prompt $J/\psi$ productions associated with a $c\bar{c}$ pair at the ILC with $\sqrt{S}=1$ TeV in Figs.(\ref{jpsiptdirect}, \ref{jpsiptprompt}). Fig.(\ref{jpsiptdirect}) shows that the color-octet and color-singlet channels have different $p_t$ behaviors. Fig.(\ref{jpsiptprompt}) shows the relative importance of the feed-down contributions from $\psi^{\prime}$ and $\chi_{cJ}$ to the direct one. The fact that the $^3P^{[8]}_J$ channel, together with $^1S^{[8]}_0$ channel, provide the dominant contributions to $J/\psi$ production cross-section can be explained by their large contributions in small $p_t$ region. More explicitly, the $^1S_0^{[8]}$ $p_t$ distribution is over the color-singlet one for $p_t<20$ GeV, while the $^3P_J^{[8]}$ distribution dominates the color-singlet one for $p_t<45$ GeV. Different from other channels which scale at least as $1/{p^6_t}$, the $^3S^{[8]}_1$ channel scales as ${1}/{p^4_t}$ due to the one gluon fragmentation mechanism. Thus even though its differential cross-section is small in low $p_t$ region, it becomes important in high $p_t$ region, i.e., $p_t>50~\textrm{GeV}$.

For our present considered production channel $\gamma\gamma\to J/\psi+c\bar{c}$, there is no $\ln(p_t/m_c)$-type large log-terms at the tree-level. This is because all the final particles are massive and the intermediate gluon should also be hard enough either to generate a $c\bar{c}$-quark pair [Figs.(\ref{FD}a,\ref{FD}b,\ref{FD}c)] or to pull the $c$ and $\bar{c}$ in different $z$ direction together to form a $J/\psi$ [Fig.(\ref{FD}d)] or to put the free $c$ (or $\bar{c}$) changing its $z$-direction to run along with the same $z$-direction of $J/\psi$ [Fig.(\ref{FD}e)]. More explicitly, for the dominant $^1S^{[8]}_0$ and $^3P^{[8]}_J$ channels, Eq.(\ref{gluonmass}) indicates that the momentum of the intermediate gluon can be small but cannot equal to zero, thus the $J/\psi$ $p_t$-distributions are under well control. This can be seen more clearly from Figs.(\ref{jpsiptdirect}, \ref{jpsiptprompt}) that the maximum differential cross-sections appear at about $p_t \sim m_c$, and it drops down for even smaller $p_t$ values; at $p_t = 0$, its cross-section is suppressed from the phase-space due to all the final particles are massive. For a higher-order calculation, there may emerge $\ln(p^2_t/m^2_c)$-type large log-terms due to extra soft parton (gluon or light quarks) coming into contribution, requiring special treatment, which however is out of the range of the present paper \footnote{For the present considered photon-photon collision processes with massive final particles, in different to the case of hadronic productions~\cite{Berger:2004cc, Watanabe:2015yca}, there is no large log-contributions from initial-state gluon showers. The gluon radiation is also suppressed from the final heavy quark lines~\cite{Berger:1993yp, Marchesini:1989yk}. Thus the higher fixed-order pQCD prediction may still be trustable in small $p_t$ region, at least for $p_t>2 m_c$. This could be similar to a NLO calculation of $e^+ e^-\to J/\psi+c\bar{c}$ at the $B$ factories~\cite{Gong:2009ng}. }.

\begin{figure}[htb]
\begin{center}
\includegraphics[width=0.48\textwidth]{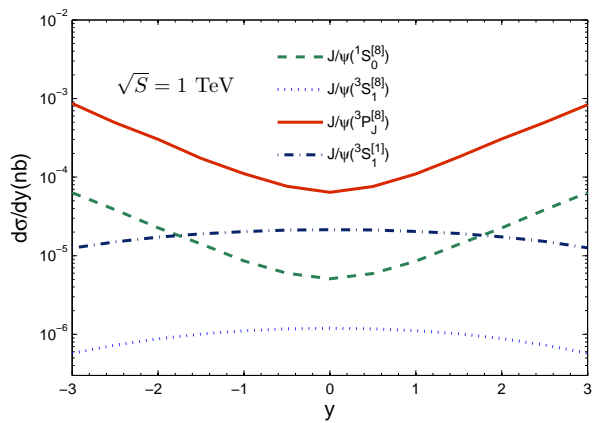}
\caption{The $y$-distributions for direct $J/\psi$ production associated with a $c\bar{c}$ pair at the ILC. The dash-dot, the dashed, the dotted and the solid lines are for $^3S^{[1]}_1$, $^1S^{[8]}_0$, $^3S^{[8]}_1$ and $^3P^{[8]}_J$, respectively. } \label{jpsiydirect}
\end{center}
\end{figure}

\begin{figure}[htb]
\begin{center}
\includegraphics[width=0.48\textwidth]{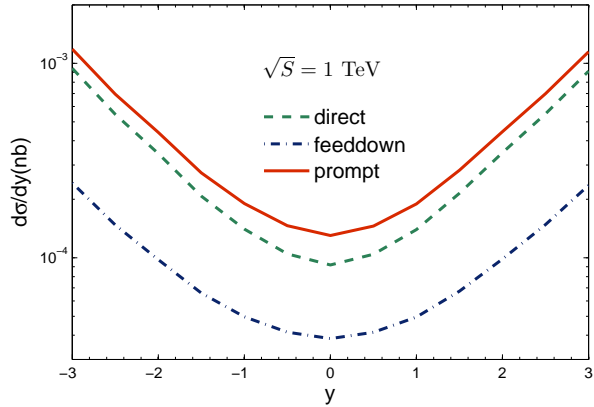}
\caption{The $y$-distributions for the prompt $J/\psi$ production associated with a $c\bar{c}$ pair at the ILC. The dashed line is for direct $J/\psi$ production, in which the contributions through the four channels have been summed up. The dash-dot line is for the feed-down contribution from $\psi^{\prime}$ and $\chi_{cJ}$ decay. The solid line is for the prompt production.} \label{jpsiyprompt}
\end{center}
\end{figure}

We present the $y$-distributions for the direct and prompt $J/\psi$ productions associated with a $c\bar{c}$ pair at the ILC with $\sqrt{S}=1$ TeV in Figs.(\ref{jpsiydirect}, \ref{jpsiyprompt}). Fig.(\ref{jpsiydirect}) shows that the rapidity distributions of the four channels can be divided into two groups, whose behavior is either concave or convex. Those two types of rapidity distributions can be adopted to distinguish the color-octet states from the color-singlet one. The concave behavior of $^1S^{[8]}_0$ and $^3P^{[8]}_J$ channels is consistent with the explanation (\ref{gluonmass}) for the large cross-section of $^1S^{[8]}_0$ and $^3P^{[8]}_J$: the largest $k^2$ is achieved at $y=0$ which leads to the lowest point of the concave, larger rapidity leads to smaller $k^2$ and larger differential cross sections.

\begin{figure}[htb]
\begin{center}
\includegraphics[width=0.48\textwidth]{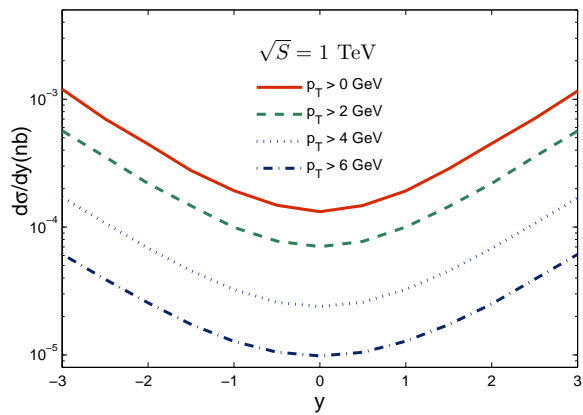}
\caption{The $y$-distributions under various $p_t$ cuts for the prompt $J/\psi$ production associated with a $c\bar{c}$ pair at the ILC. The solid, the dotted, the dashed and the dash-dot lines are for $p_t>0$ GeV, $2$ GeV, $4$ GeV, and $6$ GeV, respectively. } \label{ypromptcut}
\end{center}
\end{figure}

Considering the detectors' abilities and in order to offer experimental references, we calculate the rapidity distributions under various $p_t$ cuts. The prompt $J/\psi$ rapidity distributions for $p_t>2$ GeV, $p_t>4$ GeV and $p_t>6$ GeV are presented in Fig.(\ref{ypromptcut}). With the increment of the $p_t$ cut, up to $\sim 10$ GeV, the rapidity shapes are still dominated by $^1S^{[8]}_0$ and $^3P^{[8]}_J$ channels, following their concave behavior. The degree of the concavity decreases with with the increment of the $p_t$ cut, which is due to the fact that the term $[-(1+e^{-2y})(p_t)^2]$ in Eq.(\ref{gluonmass}) becomes more and more important for a larger $p_t$ value, providing a suppression for the cross-sections with larger rapidities. By taking a much higher $p_t$ cut, e.g. $p_t>40$ GeV, the other two direct channels, $^3S^{[1]}_1$ and $^3S^{[8]}_1$, becomes important, e.g. for small rapidity region $|y|\precsim 1.6$ their differential cross-sections are larger than those of $^1S^{[8]}_0$ and $^3P^{[8]}_J$ channels, and the prompt $J/\psi$ rapidity curve becomes a flat line for $|y|\precsim 3$.

Reviewing the long standing $J/\psi$ polarization puzzle which challenges the NRQCD theory~\cite{Affolder:2000nn}, besides the yields, we make a discussion on the polarization of prompt $J/\psi$ of the process. The polarization observable $\lambda$ for $J/\psi$ is defined as~\cite{Beneke:1998re}
\begin{eqnarray}
\lambda =\frac{d\sigma^{J/\psi}_{11}-d\sigma^{J/\psi}_{00}} {d\sigma^{J/\psi}_{11}+d\sigma^{J/\psi}_{00}},
\end{eqnarray}
where $d\sigma^{J/\psi}_{S_zS'_z}(S_z,S'_z=0,\pm1)$ is the spin density matrix, which can be calculated by FDC. The feed-down contributions from $\psi'$ and $\chi_{cJ}$ are much more involved, detailed procedures for calculating the parameter $\lambda$ for the prompt $J/\psi$ can be found in Refs.\cite{Beneke:1998re, Gong:2012ug}.

\begin{figure}[tb]
\begin{center}
\includegraphics[width=0.48\textwidth]{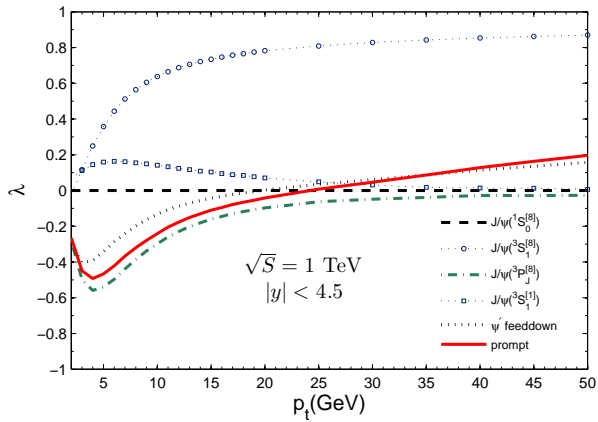}
\caption{$J/\psi$ polarization parameter $\lambda$ as a function of $p_t$. The label `prompt' represents the sum of the direct and $\psi'$ feed-down contributions. } \label{jpsipol}
\end{center}
\end{figure}

We present the polarization parameter $\lambda$ as a function of $p_t$ in Fig.(\ref{jpsipol}). As shown in Table \ref{promptcrossection}, the direct $J/\psi$ dominates the production cross-section, so in our analysis, we shall only consider the dominant feed-down contribution from $\psi'$. In fact, because the $\chi_{cJ}$ feed-down contribution to the integrated cross-section is only about $2\%$, its contribution to $\lambda$ is negligible. It is noted that the $^3S^{[1]}_1$ polarization and the prompt $J/\psi$ polarization behave quite differently from each other in whole $p_t$ region. In small $p_t$ region, the prompt polarization is dominated by $^3P^{[8]}_J$ and is longitudinal. In high $p_t$ region, it becomes transverse due to the one gluon fragmentation mechanism of $^3S^{[8]}_1$. By contrast, the polarization of the color-singlet channel is slightly transverse in small $p_t$ region, while it is almost unpolarized when $p_t>20$ GeV. Thus, besides the aforementioned $p_t$- and $y$- distributions, the polarization parameter $\lambda$ can be another useful tool to test the color-octet mechanism. \\

\section{Summary}

In the paper we have studied the photon-photon production of prompt $J/\psi$ in association with a $c\bar{c}$ pair at the future collider ILC within the framework of NRQCD. The color-octet channels, especially $^3P^{[8]}_J$ and $^1S^{[8]}_0$, provide dominate contributions to the production in small and medium $p_t$ region.

At the ILC with the $e^+ e^-$ collision energy $\sqrt{S}=1$ TeV, the color-singlet cross-section $\sigma_{^3S^{[1]}_1}$ is only $\sim 1.5\%$ of the NRQCD prompt prediction $\sigma_{\rm NRQCD}$~\footnote{Throughout the paper, we have adopted a conservative rapidity range $|y|<4.5$ to do the calculation. If taking a smaller rapidity region as $|y|<2.0$, the percentage of $\sigma_{^3S^{[1]}_1}$ to $\sigma_{\rm NRQCD}$ shall be raised up to $\sim 10\%$.}, which includes both direct and feed-down contributions. The total feed-down cross-section is sizable, providing $\sim22\%$ of $\sigma_{\rm NRQCD}$, thus those feed-down channels should be taken into consideration as a sound prediction. Moreover, the predicted $J/\psi$ $p_t$- and $y$- distributions, as well as the $J/\psi$ polarization, given by the color-singlet mechanism and NRQCD are quite different.

If taking the ILC luminosity as ${\cal L} \simeq 10^{34} {\rm cm}^{-2} {\rm s}^{-1}$, sizable $J/\psi$ events can be generated in one operation year, i.e. about $2.7 \times 10^{6}$, $2.0 \times 10^{6}$ or $1.1 \times 10^{6}$ events can be generated for $\sqrt{S}=250$ GeV, $500$ GeV or $1$ TeV, respectively. For the case of $\sqrt{S}=1$ TeV, the generated events shall be changed down to $5.0 \times 10^{5}$, $1.5 \times 10^{5}$, and $5.7 \times 10^{4}$ for $p_t>2$ GeV, $p_t>4$ GeV and $p_t>6$ GeV, respectively. All those are sizable quantities, the $J/\psi$ photon-photon production channel shall thus provide a useful platform for testing the NRQCD color-octet mechanism.

One thing need to mention is that in the paper, we do not calculate the next-to-leading order QCD correction to the channels, which might be significant. However, due to the $\alpha_s$-power suppression at higher orders, we can expect that the contributions from the color-octet channels are at least important. At the ILC, the $J/\psi$ can also be produced in associated with an open $b\bar{b}$-pair, which is out of the range of the present paper.  \\

\noindent{\bf Acknowledgments}: We thank Rong Li for helpful discussions. This work was supported in part by the Natural Science Foundation of China under Grant No.11275280 and No.11405268, and by Fundamental Research Funds for the Central Universities under Grant No.CDJZR305513.

\end{document}